\documentclass[onecolumn]{aa} 
\usepackage{graphicx}
\usepackage{txfonts}
\usepackage{color}

\usepackage{natbib}
\bibpunct{(}{)}{;}{a}{}{,} 
\usepackage{graphicx}
\usepackage{lscape}
\usepackage{rotating}

\def\simle{\lower 2pt \hbox {$\buildrel < \over {\scriptstyle \sim }$}}
\def\simge{\lower 2pt \hbox {$\buildrel > \over {\scriptstyle \sim }$}}

\begin{document}

\title{The mass function of nearby black hole candidates}

\author{Lauren\c{t}iu I. Caramete 
	\inst{1,2,}\thanks{Member of the International Max Planck Research School
(IMPRS) for Astronomy and Astrophysics at the Universities
of Bonn and Cologne}
		\and Peter L. Biermann\inst{1,} \inst{3,} \inst{4,} \inst{5,} \inst{6}
		}
\institute{MPI for Radioastronomy, Bonn, Germany
	\and
Institute for Space Sciences (ISS), Bucharest, Romania
	\and
Dept. of Phys. \& Astron., Univ. of Bonn, Germany
\and
Dept. of Phys. \& Astr., Univ. of Alabama, Tuscaloosa, AL, USA
\and
Dept. of Phys., Univ. of Alabama at Huntsville, AL, USA
\and
FZ Karlsruhe, and Phys. Dept., Univ. Karlsruhe, Germany}

%
%
%
%
%

\date{}

 \abstract
{The mass function of super-massive black holes in our cosmic neighborhood is required to understand the statistics of their activity and consequently the origin of the ultra high energy particles.}
{We determine a mass function of black hole candidates from the entire sky except for the Galactic plane.}
{Using the 2MASS catalogue as a starting point, and the well established correlation between black hole mass and the bulge of old population of stars, we derive a list of nearby black hole candidates within the redshift range $z \, < \, 0.025$, then do a further selection based on the Hubble-type, and give this as a catalogue elsewhere. The final list of black hole candidates above a mass of $M_{BH} > 3\cdot 10^{6} \, M_{\odot}$ has 5,829 entries; moreover doing a further Hubble type correction to account for the selection effects cuts down the number to 2,919 black hole candidates. Here we use this catalogue to derive the black hole mass function. We also correct for volume, so that this mass function is a volume limited distribution to redshift 0.025.}
{The differential mass function of nearby black hole candidates is a curved function, with a straight simple power-law of index -3 above $10^{8} \, M_{\odot}$, growing progressively flatter towards lower masses, turning off towards a gap below $3 \, 10^{6} \, M_{\odot}$, and then extending into the range where nuclear star clusters replace black holes. The shape of this mass function can be explained in a simple merger picture. Integrating this mass function over the redshift range, from which it has been derived, gives a total number of black holes with $z \, < \, 0.025$, and $M_{BH} > 10^{7} \, M_{\odot}$ of about $2.4 \cdot 10^{4}$, or, if we just average uniformly, 0.6 for every square degree on the sky. In some models many of these, if not all, are candidates for ultra high energy particles sources. Even if a very small fraction of the super-massive black holes produces ultra high energy cosmic rays, this should be enough to observe from the statistics of their arrivals the highly inhomogeneous distribution of the galaxies and their super-massive black holes; also this distribution may be smeared out by the possible scattering of the arrival directions of the particles by intergalactic magnetic fields.}
{}
 \keywords{Black hole physics --
                Galaxies: general --
                Acceleration of particles}
\authorrunning{L.I. Caramete and P.L. Biermann}
\titlerunning{The mass function of nearby black hole candidates}
   \maketitle
\section{Introduction}

In the quest to find the origin of ultra high energy cosmic ray particles, many proposed mechanisms use the power of super-massive black holes, either obtained from accretion, e.g. \citep{1995A&A...293..665F}, or from spin-down (e.g. \citep{1977MNRAS.179..433B}), or in some other way (e.g. \citep{2006PhLB..634..125B}). It is not a priori clear, which is the best constraining argument one should use to find out what possibly very small fraction of all super-massive black holes do produce ultra high energy particles. As long as the mechanism is not clear, we need to establish a basis for discussing all plausible super-massive black holes. Therefore it is important to obtain the mass function of black holes in order to estimate the number of black holes to within the sphere from where ultra high energy particles might come.

Since various other groups have also derived the mass function of black holes, using different methods, we have a good way to check our work.

Super-massive black holes characteristics are strongly correlated with the properties of the host galaxies, and we use this correlation to obtain abundant statistics, but also to control the quality of our data set. We wish to address a number of questions related to the super-massive black holes in the universe: a) It appears that super-massive black holes have a low cut-off in the mass distribution, and it is important to verify this and try to understand why there is such a characteristic mass; also, it appears that the black holes' mass function has an upper limit, and again we need to understand whether this is just a statistical fluke, or has a physical meaning. b) What is the  slope of the mass function? This will surely constrain our idea, on how black holes are growing. c) What is the energy input into the universe during the growth? This is an additional constraint. d) What is the number of black holes? This will constrain any search for directional correlations with ultra high energy cosmic ray events.

\section{Derivation}

Super-massive black holes are common in the centers of galaxies, and their mass is correlated with the properties of the surrounding galaxy, for instance with the velocity dispersion of the nearby stars, and with the mass of stars in the spheroidal distribution of the old stellar population.

Since the old stellar population has a spectrum peaking near 2 $\mu$ ($\nu S_{\nu}$) \citep{2006ApJ...636L..21V}, and it is this population, that correlates with black holes, we use the 2MASS catalogue as our starting point \citep{2006AJ....131.1163S}: We focus on early Hubble type galaxies \citep{2009CarameteBiermann}, and use the black hole mass spheroid correlation, e.g. \citep{1997AJ....114.1771F,1998AJ....115.2285M,1998A&A...334...87W, 1998A&A...331L...1S,2004ApJ...604L..89H,2002ApJ...574..740T,2007ApJ...665..120A} to derive the black hole mass function. We select down to 0.03 Jy at 2 $ \mu$. This is far above 2MASS catalogue completeness limit.

The procedure is as follows:

Step 1: We have first the 2 micron flux density limited sample of
10,284. We use all the distance corrections available for this
sample.

Step 2: We limited the sample to well determined Hubble types, and
only early Hubble types, and have 5,894.

Step 3: The decisive step is to eliminate all galaxies, for which
we did not have sensitivity for the presence of a black hole of a given mass equally for all Hubble types. This is a decisive step, and cuts down the sample by another factor of two, to 2,928.

For this sample of 2,928 we show the luminosity function that shows the classical Press-Schechter function \citep{1976ApJ...203..297S}, a power-law with an exponential cutoff. This agrees with what is available in the literature, but uses a different approach, and for black holes very much larger numbers.

Step 4: We get the semi-final BH candidate mass function, extending apparently down to about $10^{5}$ solar masses.

In detail then:

The mass of the central black hole is proportional to the flux density of the stellar emission at 2 $\mu$ and also proportional to the luminosity distance squared. We need to determine the proportionality constant. This constant is, however, different for different Hubble types of galaxies, and we have to determine so the constant separately, and then check for self-consistency. Since the spheroidal stellar population is much smaller in later Hubble types, these differences can be large. Because the estimate of the black hole masses uses the redshift, which is a bad indicator
 of true distance at distances less than that of the Virgo cluster, we use the available distances
 provided by the work of Barry F. Madore and Ian P. Steer (available at {\it http://nedwww.ipac.caltech.edu/level5/NED1D/intro.html}) who compiled a database of 3,065 accurate, contemporary distances to 1,073 galaxies with modest recessional velocities (that is, less than 1/8 c) published between almost exclusively 1990 and 2006.
 From this we match 429 distances in the catalog of massive black hole, and for the rest of distances less than that of the Virgo cluster we used distances with respect to the Virgo infall only provided by NED.

We use known black hole masses to calibrate the various early Hubble types; this control sample comprises 58 black holes selected from the recent literature; we use only galaxies earlier than type Sc, including type Sb. These fits give us an error bar of $\pm$ 0.4 dex, slightly worse than the $\pm$ 0.3 dex error given by more sophisticated fits using a much better but also much smaller data set \citep{2007ApJ...665..120A}. These corrections affect a large fraction of the candidate black holes. The final list above $3 \cdot 10^{6} \, M_{\odot}$ comprises 5,829 black hole candidates. The main error is $\pm$ 0.4 dex, and affects most of all the binning of the sample.

In table ~\ref{table1} we show, how many galaxies of various kinds are excluded, and it becomes obvious, that from these statistics, at worst we have a factor of less than 2 error, if all excluded galaxies of no known Hubble type turned out to be E or S0.

\begin{table}[h!b!p!]
\begin{center}
\small
\caption{}
\begin{tabular}{|c|c|c|c|}
\hline
Type&5894 selection&2928 selection&Rejected list\\
\hline
E&783&765&-\\
S0&2626&1771&-\\
Sa&879&184&-\\
Sb&1052&128&-\\
Sab&554&80&-\\
\hline
\hline
Sc&-&-&1941\\
Irr&-&-&14\\
Sy&-&-&128\\
No type&-&-&867\\
Sbrst&-&-&421\\
Unknown type&-&-&1019\\
\hline
\end{tabular}
\label{table1}
\end{center}
\end{table}

\begin{figure}[htpb]
\centering
\includegraphics[viewport=0cm 0cm 20cm 10.5cm,clip,scale=0.8]{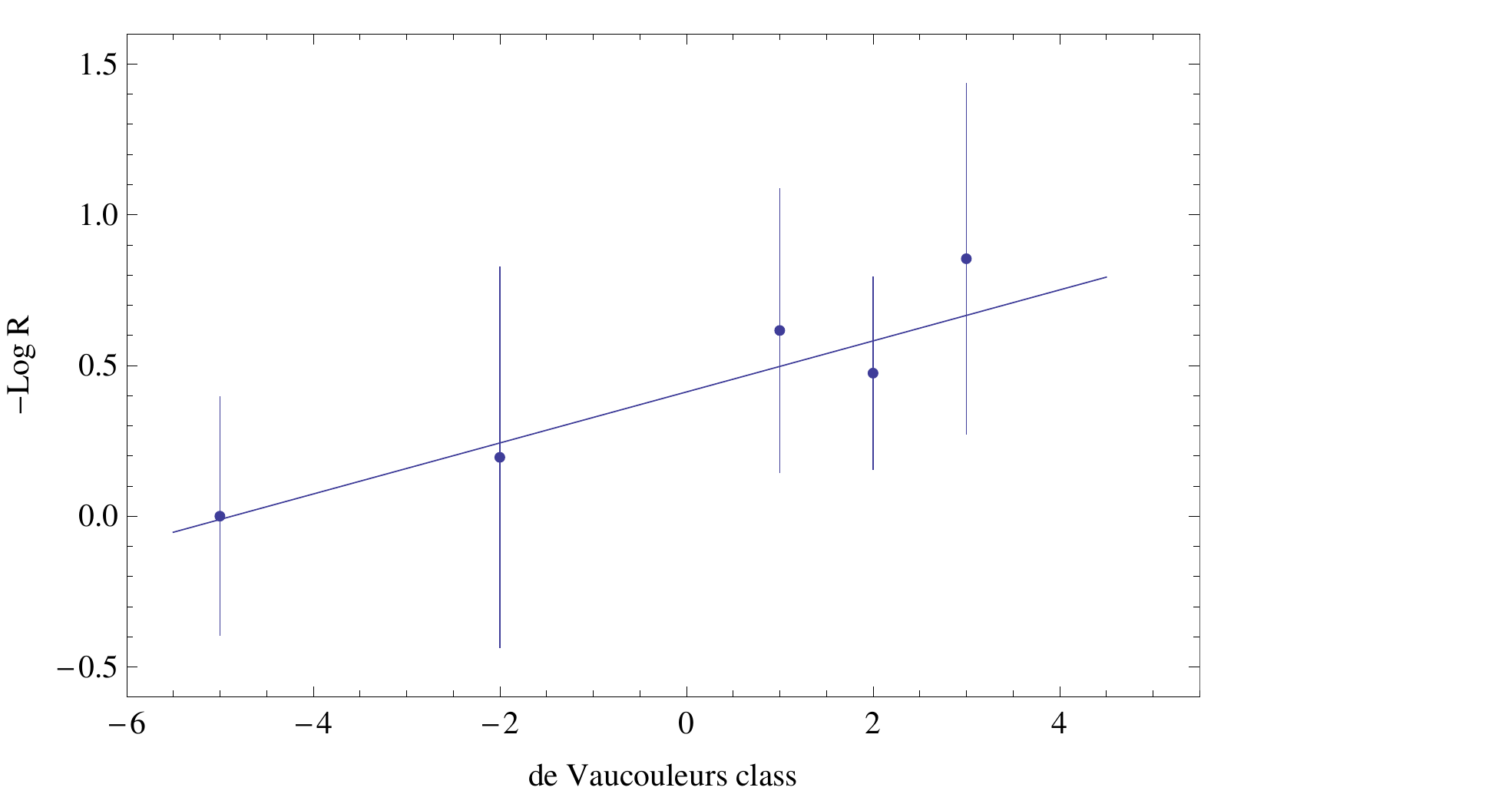}
\caption{This is the correction factor to obtain the black hole mass from the 2 $ \mu$ emission of the old stellar population, as a function of Hubble type, relative to elliptical galaxies.}
\label{BHMassCoefficientsFit}
\end{figure}

We show the correction factor dependent of the Hubble type, relative to elliptical galaxies in Fig. ~\ref{BHMassCoefficientsFit}.

\begin{figure}[htpb]
\centering
\includegraphics[viewport=0cm 0cm 20cm
12cm,clip,scale=0.7]{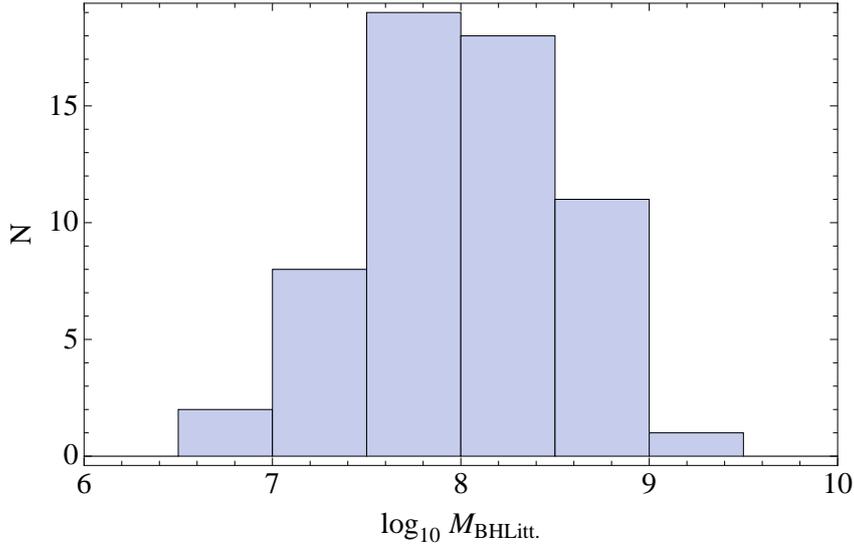}
\caption{Histogram of the calculated black hole masses for the sample
of black holes obtained from the literature.}
\label{BHHistogramOfLittMass}
\end{figure}

In histogram ~\ref{BHHistogramOfLittMass} we show that we have galaxies with black hole masses over the entire range of $10^{6} \, M_{\odot}$ to $10^{9} \, M_{\odot}$ as calibrators.

\begin{figure}[h!]
\centering
\includegraphics[viewport=0cm 0cm 20cm 10.5cm,clip,scale=0.9]{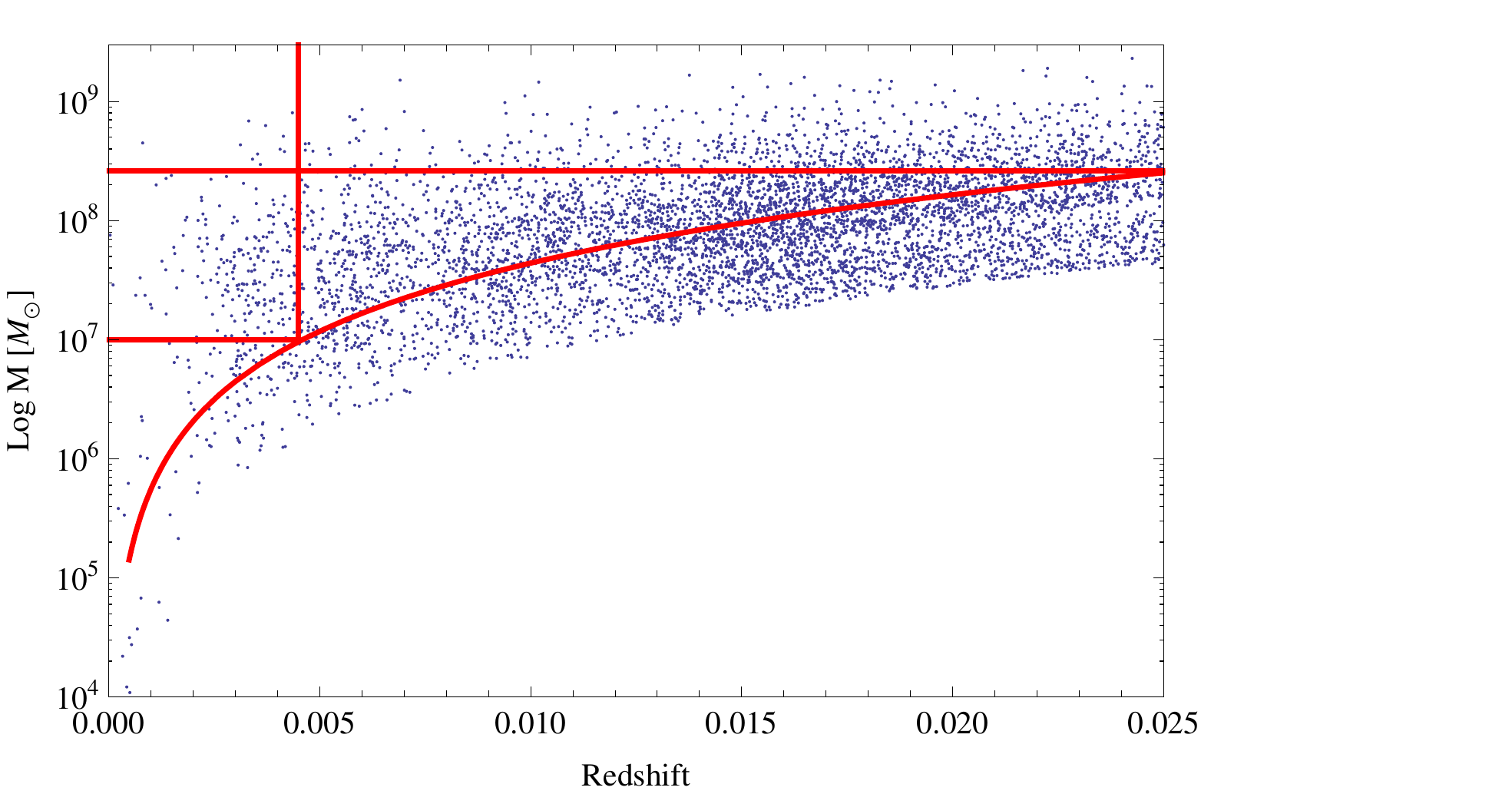} 
\caption{Plot of Mass over redshift for the massive black hole catalog
with the selection curve for elliptical galaxies in blue. This also shows the stepwise selection procedure with the two most extreme cases, the lowest mass, and the highest redshift, also in red.}
\label{histogrammassoverredshift}
\end{figure}

We derive a mass function by going to the minimum mass at each redshift, for which the list is complete to the chosen flux density in the 2MASS catalogue, starting with 0.025 and going down, always correcting for smaller volume. The sequence of steps is as follows: we obtain the number of black holes in several bins above the minimum mass we can obtain for both spirals and ellipticals, at redshift $z \, = \, 0.025$; then we go down in redshift, correcting for different accessible volume for a complete sample.  Figure ~\ref{histogrammassoverredshift} illustrates the procedure.

This diagram of black hole mass versus redshift shows clearly, that near the flux density limit, at any redshift, we have only spirals, since for a given redshift and given flux density, elliptical galaxies have an inferred black hole mass a factor of up to 5 larger than spirals, Sb spirals being the most extreme. In order to eliminate this bias, we have cut the entire sample again such that we have a black hole mass limit at each redshift, which allows both spirals and ellipticals. This procedure eliminates many, but not all of the late type spirals from our final sample to determine the mass function. The structures seen in this diagram illustrate the bias introduced by the correction factor to derive black hole masses from flux density limited data. This cuts down the fraction of spirals in the final ``Hubble-selected" sample to about twenty percent, so reducing greatly the error contribution from the Hubble-type correction shown above.

Therefore we follow the curved red line in Fig. ~\ref{histogrammassoverredshift}, get the number in that mass bin, and ratio to volume; and so we keep adding bins at lower mass until we reach the smallest bin, we wish to consider. This procedure ends at redshift $ z \, = \, 0.0045$ for $M \, > \, 10^{7} \, M_{\odot}$, and at $z \, = \, 0.0024$ for $M \, > \, 3 \cdot 10^{6} \, M_{\odot}$.

Next we show the luminosity and mass functions, using this final selection of galaxies and their black hole candidates.

\begin{figure}[htpb]
\centering
\includegraphics[viewport=0cm 0cm 20cm
12cm,clip,scale=0.8]{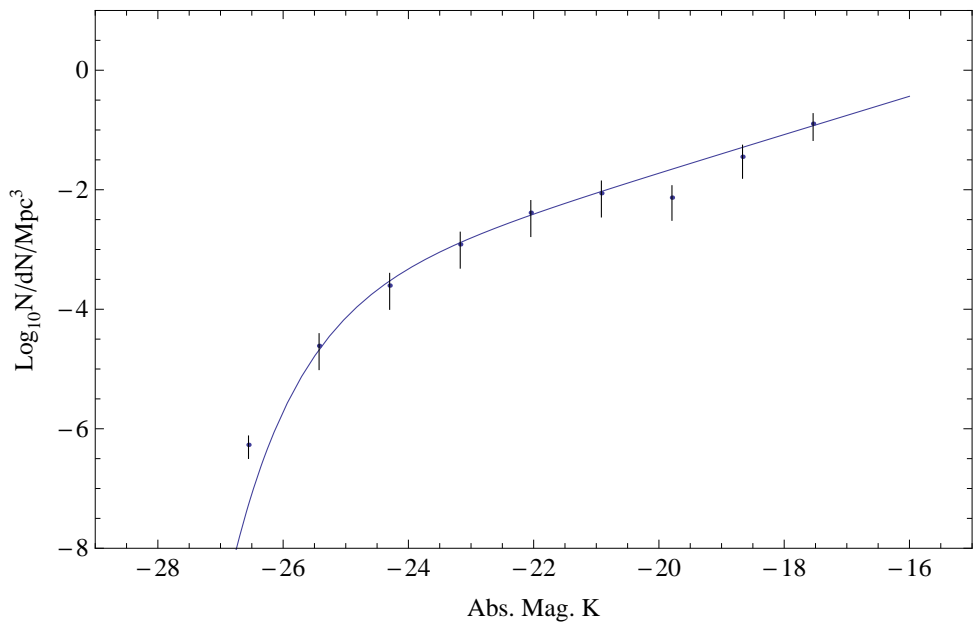}
\caption{Integral luminosity function corrected for Hubble type
sampling, 2928 objects selected in K-band, the parameters of the
Schechter function fit are: $M_{*}=-24.3, \phi_{*}=0.0008,
\alpha=-1.8$.}
\label{SchechterFunctionForBHSelectedWithFit}
\end{figure}

The luminosity functions (Fig. ~\ref{SchechterFunctionForBHSelectedWithFit}) correspond to what is in the literature, as we discuss further below; they are all well fitted with a Press-Schechter law (1974), a power-law with an exponential cutoff. Since our completeness criterion is different from what most others have used, we note that our luminosity functions are similar. Since we selected far above the completeness limit of the 2MASS sky survey, and aim to almost 80 percent at elliptical and S0-galaxies, our procedure of selection may be more complete for the task at hand, to determine the black hole mass function, independent of any other property of the black hole or the activity of its surroundings.

The errors included here include an estimate from the binning error induced by uncertain distances, and of course, also Poisson statistics; these are really relevant only for the highest mass bin, where we have 25 galaxies with their presumed black holes.

We have consciously ignored all super-massive black holes in late Hubble type galaxies; this underestimates our numbers. Correcting for the Galactic plane using a histogram of numbers per latitude bin \citep[see][]{2009CarameteBiermann} increases our number by 12.5 percent.

\begin{figure}[htpb]
\centering
\includegraphics[viewport=0cm 0cm 20cm
12cm,clip,scale=0.9]{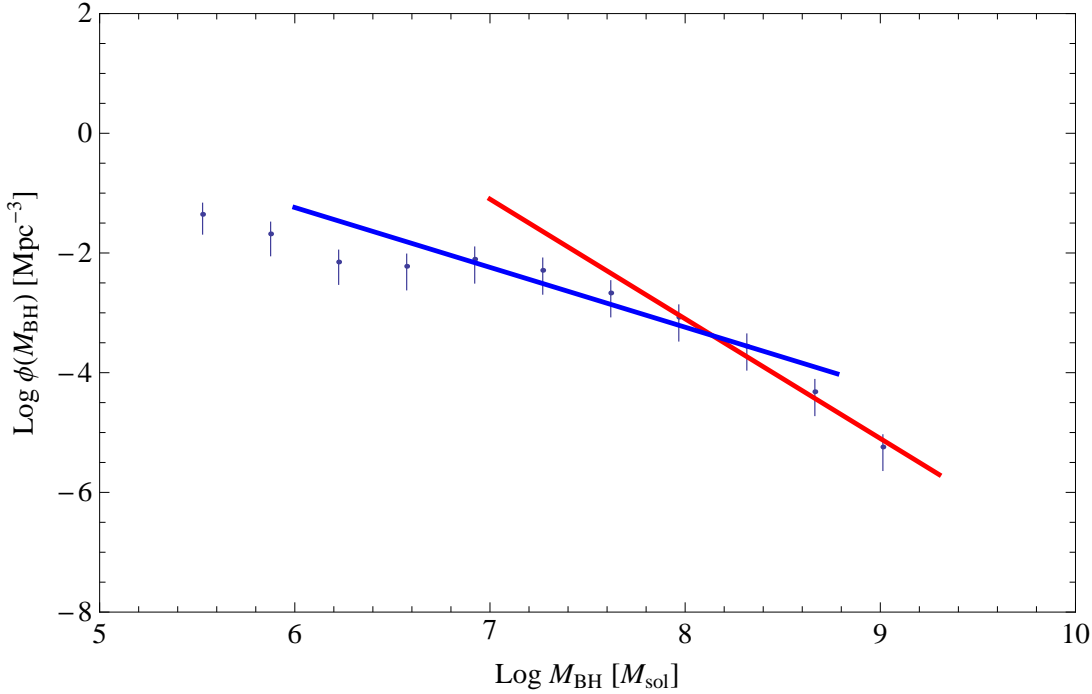}
\caption{Integral mass function corrected for Hubble type sampling,
2928 objects, the slope of the lines is: thick line $-2.0$ fitting $> \, 10^{8} \, M_{\odot}$, and thin line $-1.0$ fitting between $10^{7} \, M_{\odot}$ and $10^{8} \, M_{\odot}$.}
\label{BHMassFunctionFitCutAt310to6}
\end{figure}

We show the result in Fig. ~\ref{BHMassFunctionFitCutAt310to6}, with error bars and fits through the data.

Since it is unlikely that all these galaxies actually include a central black hole, we cut the distribution near $3 \cdot 10^{6} \; M_{\odot}$, the mass of our Galactic Center black hole; we obtain, including the correction for the missing Galactic plane (12.5 percent)

\begin{equation}
\phi(>M_{BH}) \; = \; 6 \cdot 10^{-3 \pm 0.4} \, \left(\frac{M_{BH, min}}{10^{7} \, M_{\odot}}\right)^{-1.0} \, Mpc^{-3},
\end{equation}

between about $10^{7} \, M_{\odot}$ and about $10^{8} \, M_{\odot}$, and

\begin{equation}
\phi(>M_{BH}) \; = \; 9 \cdot 10^{-4 \pm 0.4} \, \left(\frac{M_{BH, min}}{10^{8} \, M_{\odot}}\right)^{-2.0} \, Mpc^{-3},
\end{equation}

above about $10^{8} \, M_{\odot}$. These two laws are reasonably good fits, and match near $1.5 \cdot 10^{8} \, M_{\odot}$. We imposed the exponents pragmatically, since they can be determined from the data only with large error bars, strongly depending on which bins are included. Simple power-laws are adequate.

Between a nominal $ 10^{6} \; M_{\odot}$ and $3 \cdot 10^{6} \; M_{\odot}$ we have only 6 candidates, also  between $3 \cdot 10^{6} \; M_{\odot}$ and $10^{7} \; M_{\odot}$ we have 46 candidates, and between $10^{7} \; M_{\odot}$ and $3 \cdot 10^{7} \; M_{\odot}$ we already have 185. Also, in the mass range between $10^{6} \; M_{\odot}$ and $10^{7} \; M_{\odot}$ there is increasing doubt with lower masses, whether we really always have a black hole, and not just a nuclear star cluster.

Comparing this and various other fits which we have tried, we summarize these attempts including error bars from the original calibration to various Hubble types; however, after our correction for Hubble type selection there are few Sb spirals left in the sample (128). Comparing a fit through only the ellipticals and S0's and the entire sample, with a cut of low mass, and without gives us an estimate of the error in the absolute normalization of about 0.25 in the log; since there are systematics in allocating Hubble types, and including or excluding galaxies, we increase our estimate of the error by about $\sqrt{3}$ to get 0.4 in the log. We estimated the errors on the absolute scaling, and the exponent from the fit. A lowering of the low mass cutoff of a factor of 3 from $10^{7} \, M_{\odot}$ to $3 \cdot 10^{6} \, M_{\odot}$ corresponds to a negligible increase in the integral number; the integral mass function is nearly flat in that range, which of course implies that the differential mass function in this interval is consistent with zero. A final test was to cut in the mass-redshift diagram even more severely to check on edge effects in the sampling; the black hole density increased marginally, less than 0.2 in the log.

As any evolution of the black hole mass function is expected to be approximately with $(1 + z)^{4 \pm 1}$, up to a redshift of 0.025 any evolutionary effects ought to be negligible. On the other hand, it is not clear, that we have reached the distance range, over which the universe is already perfectly homogeneous: that distance might be larger than 500 Mpc (Rudnick et al. 2007, Kashlinsky et al. 2008); there is no evidence that the galaxies in the universe change their statistics drastically between 100 Mpc and 500 Mpc or even more of the local universe.

\begin{figure}[htpb]
\centering
\includegraphics[viewport=0cm 0cm 20cm
12cm,clip,scale=0.9]{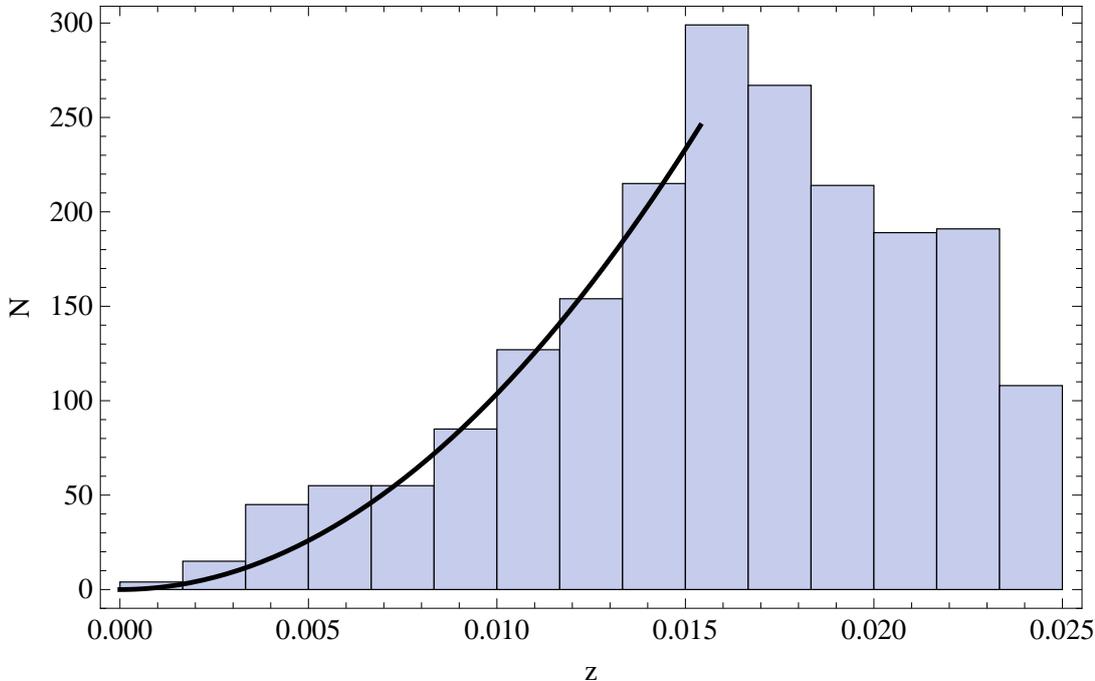}
\caption{Histograms of the black hole candidates, corrected for Hubble
type sampling, Hubble type correction, cut in BH mass at
$1*10^{8}M_{\odot}$, 2023 objects, with a $z^{2}$ fit valid for a homogeneous universe.}
\label{BHMassHistogramsAbove108}
\end{figure}

In Fig. ~\ref{BHMassHistogramsAbove108} we show our numbers of black holes above $10^{8} \, M_{\odot}$ as a function of redshift and the fit up to the redshift of a little above 0.015, where our procedure should be fairly complete for such a mass limit. A homogeneous universe fit is amazingly adequate, but the data also clearly demonstrate that near the redshift limit we are beginning to miss galaxies, here about 15 percent of the total.

\section{Comparisons}

There are a large number of explorations of the black hole mass functions, and their energy input, and here we compare with those papers.

\subsection{Numbers of black holes}

Obviously, we cannot be certain that actually all these black hole candidates are really black holes: this question was thoroughly investigated by \citet{2008ApJ...688..159G}, who could not find a large number of black holes below $10^{6} \, M_{\odot}$. What we cannot easily say, is to what degree this amounts to a selection effect. On the other hand, it has been shown by \citet{2006ApJS..165...57C}, that the correlations between galaxy parameters and the super-massive black hole apparently seamlessly merge into a correlation with nuclear star clusters. Nuclear star clusters can in turn become super-massive stars, which collapse to black holes \citep{1970ApJ...162..791S,2004Natur.428..724P}. This may in turn also explain the mass of the black holes near the transition.

The mass function is probably best determined near a mass of $10^{8} \, M_{\odot}$, where we obtain $ 9 \cdot 10^{-4 \pm 0.40} \, Mpc^{-3}$. At that mass the density is in agreement, within the error bars, with both \citet{2007ApJ...662..808L}, who show a possible range from $ 2 \cdot 10^{-4} \, Mpc^{-3}$ to $ 5 \cdot 10^{-3} \, Mpc^{-3}$, and \citet{2007MNRAS.379..841B}, who show a number near of $ 2 \cdot 10^{-3} \, Mpc^{-3}$ (correcting for Hubble constant).

\citet{2007ApJ...670...92G} have determined the black hole mass function, emphasizing the low masses, and selecting for activity: First, their numbers show, that this distribution peaks near $10^{7} \, M_{\odot}$, confirming that it may be reasonable to cut at that mass. Second, they confirm the slope which we obtain. However, their total numbers are very much lower - by a factor of about 5000 - than in \citet{2007ApJ...662..808L,2007MNRAS.379..841B}, and our results, stemming from their very different selection procedure; their selection was based on the activity of black holes.

We conclude here, that our statistics are consistent with earlier numbers.

\subsection{Energy input from black holes}

As black holes grow, they accrete from an emitting accretion disk; the luminosity depends on the growth rate, and so the final mass is a direct measure of all growth \citep{1952Zeitschr.f.NaturfA787,1973A&A....24..337S,1973blho.conf..343N}. Similarly, all well measured active galactic nuclei have radio emission, e.g., \citep{2005A&A...435..521N}, which we interpret to come from a relativistic jet, e.g., \citep{1995A&A...293..665F}; however, also in this case, all energy dissipated and emitted by the jet is a measure of growth. On the other hand, mergers of black holes, e.g. \citep{2009ApJ...697.1621G}, would more likely emit abundant gravitational waves, invisible so far in the normal universe. In order to study this we need to obtain the total growth in black holes, so their total mass. 

If we integrate from $10^{7} \, M_{\odot}$, the total mass density in black holes is given by

\begin{eqnarray}
2.8 \cdot 10^{5 \pm 0.4} \, M_{\odot} \, Mpc^{-3}
\end{eqnarray}

barely dependent on where we cut the distribution, since at low masses the integral function goes flat, and at high masses the distribution goes steep.

This compares, e.g., with the number obtained by \citet{2001ApJ...560..178K}, which has been derived from radio emission signatures, and so depends on the efficiency of the conversion process: they obtain a lower limit of

\begin{equation}
\simge \, 2.2 \cdot 10^{5} \, \epsilon_{0.1}^{-1} \, M_{\odot} \,
Mpc^{-3},
\end{equation}

where $\epsilon_{0.1}$ is the efficiency in units of ten percent. \citet{2001ApJ...560..178K} derived their numbers from a selection of higher mass black holes, and give a lower limit. One could argue that all power output by black holes in the mass range between $10^{6} \, M_{\odot}$ and $10^{7} \, M_{\odot}$ influences mostly the host galaxy, and is less relevant for the intergalactic medium; it could, for instance, mostly contribute to driving a galactic wind \citep[see][]{2008ApJ...674..258E,2008Natur.452..826B}. Above we ignored this contribution.

Another way of looking at this is by simply converting the units, and then the mass density derived from this is

\begin{equation}
1.9 \cdot 10^{-35 \pm 0.40} \, g \, cm^{-3}
\end{equation}

which corresponds to a cosmological density of (using for the Hubble constant 70.1 km/s/Mpc, the critical density is $9.77 \cdot 10^{-30}$; \citet{2009ApJS..180..330K}

\begin{equation}
\Omega_{BH} \, = \; 2 \cdot 10^{-6 \pm 0.40}
\end{equation}

where the error bar in the first factor corresponds to a range of $0.8 \cdot 10^{-6}$ to $5 \cdot 10^{-6}$.

Assuming, that growth of black holes is mostly via baryonic accretion \citep{1973A&A....24..337S,1973blho.conf..343N}, the energy input into intergalactic space from the production of all these black holes is given by

\begin{equation}
1.9 \cdot 10^{-15 \pm 0.40} \, \epsilon_{0.1} \, erg \, cm^{-3},
\end{equation}

which is by a factor of order $4.5 \cdot 10^{-3}$ - excluding the uncertainties - far below that due to the microwave background (at $4.2 \cdot 10^{-13} \, erg \, cm^{-3}$); this number is an average over all space, derived from the local universe to redshift 0.025.

We need to compare this estimate with the energy density visible directly in electromagnetic background, with the energy density inferred to be there in the form of turbulent motion, heat, magnetic fields, energetic particles, as well as comprehensive simulations of all such processes.

The electromagnetic radiation background in other wavelengths is approximately (compilation by Kneiske (2009), private communication): at FIR and optical frequencies lower than the microwave background by $5 \cdot 10^{-3}$, in X-rays down by $5 \cdot 10^{-5}$, decreasing with higher frequencies then to a factor of $1.5 \cdot 10^{-6}$ down for $\gamma$-rays. The energy input from the growth of black holes, even at only 10 percent efficiency as used above, exceeds all other energy densities, and could be marginally consistent with the FIR and optical background.

The energy input has also been estimated by \citet{2001ApJ...560L.115G}, and \citet{1998A&A...333L..47E}. En{\ss}lin et al. obtain $(2.5 - 4.0) \cdot 10^{-16} erg/cm^{3}$, while Gopal-Krishna et al. get $2 \cdot 10^{-15} erg/cm^{3}$ in filaments, similar to what we obtain. We conclude that our numbers are consistent with earlier estimates, derived differently.

We will compare with the very uncertain energy densities of ultra high energy particles elsewhere, since such a comparison strongly depends on the spectrum of energetic particles below the energies at which we can observe them.

We should also compare this with the numbers derived from simulating the turbulent dynamo in the intergalactic medium \citep{2008Sci...320..909R}; Ryu \& Kang (2009), private communication: in this case we need to compare with the energy input into thermal motions and magnetic fields: They obtain a magnetic field energy density of $2 \cdot 10^{-17} \, erg/cm^{3}$ averaged over the universe at zero redshift; in filaments a magnetic field energy density of $1 \cdot 10^{-16} \, erg/cm^{3}$, in average a thermal energy density $3 \cdot 10^{-16} \, erg/cm^{3}$, and in kinetic energy a density $1 \cdot 10^{-15} \, erg/cm^{3}$. All this energy in these simulations derives from the gravitational wells of the dark matter large scale structure, including the galactic halos. To inject far more energy from the activity and growth of central black holes could present a serious problem. The energy input from active galactic nuclei into magnetic fields might be far less than 10 percent efficient; to satisfy observational constraints we have the condition then

\begin{equation}
1 \cdot 10^{-2 \pm 0.4} > \epsilon_{B}
\end{equation}

where $\epsilon_{B}$ is here the efficiency to channel energy from black hole growth into overall magnetic field. We might satisfy this by arguing that much less than 10 percent of all black holes ever become radio galaxies, or alternatively that all super-massive black holes are radio galaxies for much less than 10 percent of their time, and much less than 1 percent of their accretion energy is channelled into magnetic fields (we need a factor of $2 \cdot 10^{-2}$). This may then be consistent with the reasoning of \citet{2001ApJ...560..178K}, who use just the manifest energy blown into the intergalactic medium by radio galaxies, traceable through their nonthermal radio emission, to derive numbers for the black hole density.

There are several ways out of this possible dilemma: an obvious one is that the fraction of energy going into various channels, electromagnetic, kinetic, magnetic fields is far from settled. A last resort is to note that much of the energy emitted during the growth of black holes via mergers could go into gravitational waves.

\subsection{Extrapolation to lower masses?}

Now, we have to remember that we cannot be absolutely sure that the low mass cutoff is in fact at $10^{7} \, M_{\odot}$, it might be at stellar scale, e.g., \citep{2002A&A...392..909C,2009AAS...21343707I}. There are very few confirmed black hole candidates with a mass, that significantly exceeds 10 $M_{\odot}$, and is below $10^{5} \; M_{\odot}$: \citep[see][]{1996A&A...314..521V,2004ApJ...616..821T,2005ApJ...634.1093G,2006MNRAS.370L...6P,2006A&G....47f..29M,2007Natur.449..872O,2007ApJ...661L.151U, 2008MNRAS.387.1707C,2008ApJ...678L..17S,2009ARep...53..232A,2009Natur.460...73F}. One of the most convincing cases for an intermediate mass black hole is the candidate in the cluster G1, galaxy M31 also known as Andromeda.

Since the integral mass function flattens out just near the mass, where there may be very few black holes anymore, below $3 \cdot 10^{6} \, M_{\odot}$, there maybe little mass there. The mass to there is only $2.8 \, 10^{5 \pm 0.4} \, M_{\odot} \, Mpc^{-3}$. This is far below the numbers for stellar mass black holes, according to some estimates \citep{2002MNRAS.334..553A}; it is intriguing to note that the total mass in stellar size black holes inside our Galaxy is estimated to be of order a few $10^{9} \; M_{\odot}$; our law clearly does not connect to this. It is then reassuring, that there is little evidence for black holes between 20 and $ 10^{5} \, M_{\odot}$ \citep{2004ApJ...607...90B,2005ApJ...619L.151B,2008ApJ...683L.119B,2008AJ....136.1179B,2009ApJ...690.1031B,2006NewAR..50..739G,2007ApJ...670...92G,2007ASPC..373...33G,2008ApJ...688..159G}, and it appears, that there are many black holes between $10^{6} \, M_{\odot}$ and $10^{7} \, M_{\odot}$, but probably not enough to fill the distribution. These stellar black holes add a factor to the cosmological density of black holes overall.

Here we add some more arguments to the case that filling this hole in the black hole mass distribution is fraught with challenges:

The finding of few black holes in this gap region is consistent with many other arguments \citep{2002A&A...382L..13K}, saying that there may be no or very few intermediate mass black holes. On the other hand, since the accretion statistics of any hypothetical intermediate mass black holes are not known, the observed distribution of candidate sources \citep{2002A&A...382L..13K} might just be compatible with our statistics. If we were to require that this energy density stays at the level of the $10^{-5}$ fluctuations in the microwave background, then $M_{BH, min} $ would have to be beyond the upper end of the distribution. So whatever energy gets transmitted during the growth of black holes to the observed distribution, it cannot have much of any effect on the low wave-number fluctuations in the microwave background.

To spin this out, one could imagine that black holes grow initially fast by feeding on dark matter \citep{2006A&A...458L...9M,2005A&A...436..805M}, and then continue to grow mostly by mergers; in such a scheme most of the black hole feeding would have come from dark matter. The gravitational waves emitted by all these growth events and mergers might obey the equation of state for dark energy \citep[see][]{2008arXiv0811.4484M}. This does not look like a convincing concept, in which we have first all dark matter to be black holes  \citet[see][]{2009ApJ...705..659A}, and second all dark energy gravitational waves.
At present this extreme speculation does not appear to be a viable solution.

\section{The slope of the mass function}

The shape of the mass function is a constraint on the growth of the distribution.  A simple powerlaw at high masses suggests a self-similar process.

\citep{1979ApJ...229..242S} have shown how to estimate the building of a mass function from a repeated merger process, building on yet earlier work. Here we wish to ask, whether their approach can also explain the slope which we find here.

We assume that all black holes grow by merging with other black holes, and that baryonic accretion is just a multiplying factor, like perhaps always adding a factor of 1.5 in mass. It may well be a narrow distribution of fractional mass added.

Then writing the coalescence rate as $\sim \, (Mass)^{\lambda}$ they find
that the resulting mass function in its power law part becomes $\sim m^{-3 \lambda /2}$. In our context this implies that $\lambda \, = \, 2.0$. This then says that the product of cross section, typical velocity, and sticking probability run with this combined dependence on mass. We argued in \citet{2009ApJ...697.1621G}, that the cross section runs as $(Mass)^{1/2}$. On the other hand, small systems like the local group, when the earlier mergers occur, have a lower typical velocity of order 200 km/s than larger systems like clusters of galaxies, with velocity dispersions of order 2000 km/s. This leads to a crude dependence estimate of $(Mass)^{+1}$, with a large uncertainty. The stickiness we estimate at 100 percent for the cross-section chosen, given a few spiral-down orbits. The combination would imply $\lambda \, = \, 1.5$, compatible with what we obtain, allowing for large error bars. It seems possible that the merger cross section rises more steeply with mass than what we assumed here, and an exponent close to unity would allow a better match with data. We conclude that mergers between black holes might be able to explain the entire mass distribution.

However, we have to ask whether such a growth process could operate in a very similar way for nuclear star clusters, as for black holes. All such merger arguments may work as well for nuclear star clusters as for black holes surrounded by stars.

In such a picture the upper end of the distribution is just the maximum that can be reached given the density of galaxies, and the mass of the central black holes.

\section{The transition in mass}

Why is there a minimum mass in super-massive black holes?

As we argued earlier, the data and much work by \citet{2008ApJ...688..159G} and others show, that there are very few black holes near to and below $10^{6} \, M_{\odot}$. There is a variety of exploratory ideas why this is so \citep{2004Natur.428..724P,2006A&A...458L...9M,2005A&A...436..805M}. It seems plausible to assume, that the transition from massive black holes to nuclear star clusters holds a clue to solving this question. A possible transition from a nuclear star cluster to a super-massive star has been discussed by \citet{1970ApJ...162..791S}, and again by \citet{2004Natur.428..724P}, with the latter team arguing for a transition to a super-massive black hole \citep{1972AAa,1972AAb,2003ApJ...591..288H}. However, the results of \citep{1972AAa,1972AAb} preclude any contribution from super-massive stars near or above $10^{6} \, M_{\odot}$, since such stars explode completely due to an instability given by General Relativity, leaving no black hole behind, we do need a mechanism that manages to give black holes right below this cutoff. \citet{2008A&A...477..223Y} have shown that wind mass loss effectively competes with agglomeration, and so would limit massive stars of a few hundred $M_{\odot}$ to below 100 $M_{\odot}$; this implies that it would be difficult to get past this barrier in mass. On the other hand, agglomeration is a runaway process, while stellar winds are a quasi-steady process, basically limited to the Eddington luminosity; so perhaps more extreme conditions are required to get a serious run-away in agglomeration.

There are a number of processes, that contribute: One is the simple momentum exchange between stars, with a time scale of \citep{1942psd..book.....C,1962pfig.book.....S,1987degc.book.....S,1987gady.book.....B}:

\begin{equation}
\tau_{grav} \, = \, \frac{\sigma_{\star}^{3}}{8 \pi G_N^{2} m_{\star}^2 n_{\star} \Lambda}
\end{equation}

where $\sigma_{\star}$ is the velocity dispersion of the stars in the system, assumed to be in virial equilibrium, $m_{\star}$ is the mass of the stars,
$n_{\star}$ is the density of the stars, and $\Lambda$ is the Coulomb logarithm, typically of value 20. Considering nuclear star clusters it is hard to see why this process by itself would lead to a sudden transition at a specific mass, although a gravo-thermal catastrophe could in principle do this \citep{1987degc.book.....S}; however, this process by itself would suggest, that the lower masses become a black hole, and the higher masses remain a star cluster, contrary to observation.

On the other hand, the agglomeration of stars is governed by their collision time scale, which is

\begin{equation}
\tau_{aggl} \, = \, \frac{1}{ N_{\star} n_{\star} \sigma_{\star} \Sigma_{\star}}
\end{equation}

where $N_{\star}$ is the total number of stars, $n_{\star}$ is the density of stars, and $\Sigma_{\star}$ is the cross section of typical stars. We need only one star to start a runaway coalescence, and that is why also have the factor $N_{\star}$. The question is whether either of these two processes or a combination of the two would allow for a transition at a specific mass of a nuclear star cluster such, that a short range of masses is pinpointed. \citet{2007ASPC..367..697M} suggest, that there is a very long mass range, in which the process of agglomeration can give a large variety of masses, intermediate mass black holes. Therefore their conclusion is that this would not lead to a relatively sharp transition.

Some have suggested \citep{2008ApJ...677..146P}, that in the case of a binary black hole merger of equal masses a gravitational rocket effect could eject black holes from galactic centers, and one could so imagine, that all lower mass black holes might form, but no longer be in galactic centers. In such a speculation these black holes below the transition point would exist, but be invisible. However, in \citet{2009ApJ...697.1621G} we show that this process is unlikely to be statistically relevant.

In summary, accepting the process of agglomeration, what could modify the conclusion of previous authors, that a variety of masses is formed, and in contrast narrow down the mass range for the transition mass?

There are several avenues to consider:

First, galaxies grow by merging, starting from some minimum size: Could this minimum size of a central black hole correspond to the minimum size of a galaxy? That is hard to maintain, even considering, that \citet{2007ApJ...663..948G} have identified a minimum mass of order $5 \cdot 10^{7} \, M_{\odot}$, most of it in dark matter. To grow a galaxy like ours, with a central black hole close to the low mass cut-off, would require so many merger events, that it is hard to see that much of any connection to the minimum galaxy could survive with a signature, except for properties that survive in all galaxies, independent of whether they contain a black hole at their center.

Second, in a merger process we do obtain a central spike in dark matter from the merger density profile \citep{1997ApJ...490..493N,1998ApJ...502...48K,1998ApJ...499L...5M}:
$
\rho_{dm} \, \sim \, (x^{\gamma_{sp}} ( 1+ x)^{2})^{-1}
$
where $x = r/r_c$, a scaled radial coordinate, and $\gamma_{sp}$ is of order unity; various variants of this formula have been discussed \citep{1998ApJ...502...48K}. This implies a central dark matter component with a mass enclosed with $R$ of $M_{R, dm} \, \sim \, R^{3 - \gamma_{sp}}$. Applying this first just to stars as well implies a radial dependence of $n_{\star} \, \sim \, 1/x^{\gamma_{sp}}$, of velocity dispersion of $\sigma_{\star} \, \sim \, x^{(2 - \gamma_{sp})/2}$. The combination yields to $\tau_{aggl} \, \sim \, x^{ (3 \gamma_{sp} -1)/2}$, so to an arbitrarily short time scale of stellar agglomeration $\tau_{aggl}$ at the center of merged galaxies. This would then suggest, that all galaxies should have a central super-massive black hole, and not just those above a specific mass; this is again in contradiction to data. However, including the process of massive star formation, e.g.,\citep{2009ApJ...697.1741B} near the center of a galaxy might require a certain minimum amount of gaseous turn-over by star (star formation, mass ejection by winds, and explosions), and so it is conceivable that these processes define a threshold for run-away agglomeration.

Third, in a speculation on the nature of the dark matter particle's, \citet{2006A&A...458L...9M,2005A&A...436..805M}, following earlier work, have suggested that a keV Fermion would naturally identify a mass scale from a degenerate configuration, and this mass for a dark matter particle is consistent with a few $10^{6} \, M_{\odot}$. This mass appears consistent with a) the low mass cutoff in the galaxy distribution found by \citet{2007ApJ...663..948G}, b) the early star formation \citep{2006PhRvL..96i1301B,2007ApJ...654..290S,2009ApJ...700..426L}, and c) new galaxy data interpretation \citep{2009arXiv0907.0006D}. For some of these applications it is important to remember, that a thermal keV particle as a dark matter candidate already is excluded \citep{2007astro.ph..3673S}; it would have to be substantially sub-thermal.

Fourth, if supermassive stars at all such masses do form, evolve, and explode, it is conceivable, that they do not form a black hole, but
a naked singularity \citep{2007Prama..69..119J}, which then due to quantum loop gravity effects disperses, leaving no remnant behind. This would have to happen just in the mass range between about 20 and $\sim \, 10^{6} \, M_{\odot}$, but neither above or below.

Fifth, in another concept, collapsing stars might undergo a phase transition to a SUSY-state \citep{2006HEDP....2...97C} before forming a black hole, which then also blows up all the star, leaving no remnant behind.

Finally, we need to remember, that in agglomeration of stars any massive star near to or above $10^{6} \, M_{\odot}$ becomes unstable and explodes, and does so on very short time scales, leaving no remnant behind \citep{1972AAa,1972AAb}. Therefore, all stars reaching to just below this limit will become a black hole, and perhaps with the abundances of the early universe, those very early massive stars near to but below instability threshold will also become black holes rather than completely explode. This instability defines a threshold, and makes it obvious, why the initial black holes - in the agglomeration picture - cannot be from above about $10^{6} \, M_{\odot}$. If an evolving star moves up the main sequence, steadily growing in mass, it gets ever more unstable simply due to the increasing fraction of pressure held by radiation. When it approaches the Appenzeller \& Fricke instability, it might just collapse, forming always a black hole just below threshold. A proper evolutionary calculation is required to test this.

One could imagine, that such a dark matter core would help induce a runaway agglomeration process among stars as well; both run-away agglomeration, and degenerate dark matter cores, or even a combination of both, seem feasible.

In conclusion, we have no definitive solution, why a relatively sharp transition is found in data; run-away agglomeration, perhaps helped along by other processes, seems like a good concept to explore further.

\section{Sources and transport of ultra high energy particles}

Finally, we need to know how many black holes are out there, possibly contributing to the production of ultra high energy particles. In many models the power of black holes scales with mass \citep{1995A&A...293..665F}, so that we will focus on the higher masses.

Our mass function then extrapolates within the maximum redshift used, 0.025, or approximately 100 Mpc, to

\begin{equation}
2.4 \cdot 10^{4 \pm 0.40}
\end{equation}

candidates, so on the sky 0.6 per degree squared; however, note that the spatial distribution is highly inhomogeneous. Going to only those black holes above $10^{8} \, M_{\odot}$, reduces this number to one per about 10 degree squared, so an area of about 1.8 degree radius, somewhat less than the correlation radius determined by the correlation analysis of Auger \citep{2007Sci318938T,2008APh....29..188P}. To reduce the number of black holes just using the criterion of mass to a sample size of a few hundred as used by Auger, implies a mass threshold in the distribution of about $3.5 \cdot 10^{8} \, M_{\odot}$, firmly eliminating the radio galaxy Cen A \citep{2009MNRAS.394..660C}, since it has only $(5.5 \pm 1.0) \cdot 10^{7} \, M_{\odot}$. A second criterion is required to provide a really narrow selection, such as strong radio emission could be \citep[see][]{2009NuPhS.190...61B}, as indicative of clear particle acceleration.

However, since most black holes probably are active in the spin-down limit \citep{1977MNRAS.179..433B}, we can adjust the numbers some more: The lifetime of activity in the spin-down limit is very long, and so the power emitted in the form of a jet, and radiation, is only a small fraction of the Eddington limit, of order percent or even less. This entails for protons, that only black holes above $10^{9} \, M_{\odot}$ can produce particles to $10^{21} \, {\rm eV}$. If we allow heavy nuclei with charge of order 10, then black holes above $10^{7} \, M_{\odot}$ are allowed, but we need then either an ongoing starburst to provide the particles for injection near PeV, or a latent such particle population left behind at substantial energy density to allow injection; this latter option appears very difficult, since cosmic rays leak on rather short time scales, especially at high energy such as PeV. Therefore of all the many black hole candidates $ \simeq 10^{7} \, M_{\odot}$ discussed here, only those with a starburst and an AGN at the same time are good candidates; otherwise only those few $ \simeq 10^{9} \, M_{\odot}$ and radio galaxy signatures are candidates to produce ultra high energy cosmic rays.

Since galaxies, and especially early Hubble type galaxies, are highly clustered on the sky, and in the universe, any subset of super-massive black holes presumed to be candidates of ultra high energy cosmic rays will probably be even more clustered; this spatial distribution might show up in the arrival direction statistics.

\section{Conclusions}

We have derived the mass function of black hole candidates in the centers of galaxies down to $3 \cdot 10^{6} \, M_{\odot}$. At high mass it is consistent with a straight steep power-law, of slope -2 in the integral mass function. In fact, since all and everyone of the black holes, which we could check on, has been declared a black hole by more sophisticated measurements, we feel fairly confident, that almost all the black hole candidates are in fact black holes; the most likely problem with such an identification is below the considered mass range.

First, we conclude that this distribution function cannot possibly continue through the gap between about 30 and $10^{6}\, M_{\odot}$, consistent with \citet{2006NewAR..50..739G,2007Natur.445..183M,2008MNRAS.389..379M}. Coming from the higher masses it flattens near $10^{7} \, M_{\odot}$, giving a contribution to the cosmological density of only

\begin{equation}
\Omega_{BH} \, = \; 2 \cdot 10^{-6 \pm 0.40}
\end{equation}

consistent with many other earlier estimates. Like in some other analyses our distribution begins to flatten already below $10^{8} \, M_{\odot}$ \citep{2007ApJ...662..808L,2007MNRAS.379..841B}.

As a corollary we note, that \citet{2009arXiv0905.2535F} has pointed out that black holes constitute a huge reservoir of entropy: Integrating the mass function to obtain the total entropy we find here shows that the entropy is dominated by the biggest black holes, and is quite large. However, allowing for the errors in the exponent, the entropy might be logarithmically evenly distributed among all masses. Therefore the entropy is essentially unaffected by the speculation, that the black hole mass function may extend to lower masses, and in fact, it is curious that a logarithmically even distribution in entropy seems probable to be realized in Nature. However, this implies that the summed entropy of all super-massive black holes is a reasonably well defined quantity, a new cosmological parameter, that requires an understanding.

As a second corollary we note that the early energy input of gravitational waves from the first merging black holes approaches oddly the energy density of dark energy at a very high redshift, of order 50, possible since very massive stars may have formed at even earlier redshift, such as 80 \citep{2006PhRvL..96i1301B}.

Second, the energy input from the growth of all these black holes is at the fraction of a percent level of the microwave background, which may mostly be lost in expansion, but some traces should remain. If some of this energy is transferred to magnetic fields, beyond what is given to intergalactic magnetic field from the large scale gravitational potential wells, the effect on the scattering of ultra high energy particles could be strong. In fact, this channel might present a problem. The fraction of black hole growth energy going into this channel is constrained to be less than a few percent.

Third and last, just going to a distance of 100 Mpc, or redshift of 0.025, and going to $10^{7} \; M_{\odot}$ our results give $2.4 \cdot 10^{4}$ black holes, or on the sky 0.6 per every square degree. There is some evidence that every super-massive black hole is active at an observable level in radio emission \citep{2005A&A...435..521N}. Radio emission implies non-thermal particles, and so the speculation, that each and everyone might contribute also to ultra high energy cosmic rays, is not immediately discountable. More specific models, with testable predictions, are required to investigate this possibility. For heavy nuclei as ultra high energy cosmic rays the black holes would require evidence of both a star-burst and nuclear activity, such as in Cen A; for protons only very massive black holes are allowed, such as M87. Evidence of a particle accelerator in the form of a radio galaxy limits the choice severely.

Allowing for all such possibilities, a priori, there are many candidates every square degree on the sky: the distribution of these black hole candidates is highly clustered, and this clustering may survive the magnetic scattering by intergalactic magnetic fields to show up in the studies of the arrival directions of ultra high energy cosmic ray particles \citep{2008ApJ...682...29D,2008Sci...320..909R}.

\section{Acknowledgements}

Useful and inspiring discussions with R. Beck, J. Becker, S. Britzen, L. Clavelli, T. En{\ss}lin, L. Gergely, Gopal-Krishna, B. Harms, E.van den Heuvel, J. Irwin, P. Joshi, H. Kang, T. Kneiske, P.P. Kronberg, N. Langer, T. Maccarone, S. Portegies-Zwart, J. Rachen, D. Ryu, N. Sanchez, T. Stanev, H. de Vega, and P. Wiita are gratefully acknowledged.
Work with PLB was supported by contract AUGER 05 CU 5PD 1/2 via DESY/BMB and by VIHKOS via FZ Karlsruhe; by Erasmus/Sokrates EU-contracts with the universities in Bucharest, Cluj-Napoca, Budapest, Szeged, Cracow, and Ljubljana; by the DFG, the DAAD and the Humboldt Foundation; and by research foundations in Japan, Korea, China, Australia, India, Europe, the USA and Brazil. LIC is partially supported by CNCSIS Contract 539/2009 and CNMP Contract 82077/2008. LIC wishes to express his thanks to the MPIfR for support when finishing this project. This research has made use of the NASA/IPAC Extragalactic Database (NED) which is operated by the Jet Propulsion Laboratory, California Institute of Technology, under contract with the National Aeronautics and Space Administration. This research also made use of the ViZier system at the Centre de Donne{\'e}s astronomiques de Strasbourg (CDS) (Ochsenbein et al. 2000).

\bibliographystyle{aa} 
\bibliography{references} 

\end{document}